\documentclass[conference]{IEEEtran}

\usepackage{times}
\usepackage{graphicx}
\usepackage{multirow}
\usepackage{subfigure}
\usepackage{url}
\usepackage{amssymb,amsmath}
\usepackage{algorithm}
\usepackage{algpseudocode}
\usepackage{dblfloatfix}
\usepackage{todonotes}

\IEEEoverridecommandlockouts

\newif\ifpaper

\papertrue

\ifpaper
\title{Application of Information Centric Networking to NoSQL Databases: the Spatio-Temporal use case}
\else
\title{Application of Information Centric Networking to Spatio-Temporal Databases - Extended Version}
\fi

\author{\IEEEauthorblockN{Andrea Detti, Michele Orru, Riccardo Paolillo, Giulio Rossi, Pierpaolo Loreti, Lorenzo Bracciale, Nicola Blefari Melazzi}
	\IEEEauthorblockA{Electronic Engineering Dept., University of Rome "Tor Vergata", Italy\\ Consorzio Nazionale Interuniversitario per le Telecomunicazioni (CNIT) \\ 
		Email: name.surname@uniroma2.it}
	\vspace{-10pt}
	\thanks{This research was partly funded by the EU H2020 Bonvoyage and EU-JP H2020 ICN2020 projects}
	}

\begin{document}

\maketitle

\begin{abstract}
This paper explores methodologies, advantages and challenges related to the use of the Information Centric Network technology for developing NoSQL distributed databases, which are expected to play a central role in the forthcoming IoT and BigData era. ICN services make possible to simplify the development of the  database software, improve performance, and provide data-level access control. We use our findings for devising a NoSQL spatio-temporal database, named OpenGeoBase, and evaluate its performance with a real data set related to Intelligent Transport System applications.

\end{abstract}



\section{Introduction}

In the forthcoming era of IoT and Big Data, NoSQL database technologies are expected to play a central role for the information management, due to their  ability to support large volumes of read-write operations and to be easily distributed on different servers. NoSQL databases store generic objects such as JSON ones. Each object contains all related data, strongly simplifying the operations to achieve data consistency in distributed environment. 


When load increases, the resources of a NoSQL database can be scaled "horizontally" by adding new servers, thus forming a database cluster that is exposed to client applications as a single entity. Clients interact with one or more front-end servers, which in turn contact back-end servers that satisfy front-end requests exploiting local storage spaces. Typical procedures carried out by front-end servers are routing of NoSQL operations (queries, insertions, deletions) toward back-end servers, access control, aggregation and post processing of results. The overall architecture can include other kind of servers to perform specific tasks, such as system configuration, security operations, global indexing etc. 

Usually, the communications between front-end and back-end servers are handled by a TCP/IP network, which sets up connections between these entities to push or pull data. However, we argue that the emerging Information Centric Networks (ICNs) \cite{jacobson2009networking} can be effectively applied for these communications. An ICN is based on a new network layer, designed to provide users with \textit{named objects}, rather than end-to-end connections. A named object is a bundle of data, with a limited size of few kB, uniquely identified by a hierarchical name. To some extent, the ICN services resemble those of a Content Delivery Network, but with a finer, packet-level, granularity.




An ICN provides: a \textit{secure name-based Application Programming Interface} (API), for requesting objects rather than connections; \textit{routing-by-name}, for forwarding requests towards data sources on the base of what (object name) is requested rather than where the request should go (destination IP address); \textit{in-network caching}, for moving popular objects on the network edge thus reducing response time and server load; \textit{multicasting}, for reducing network traffic and server load in case of concurrent data requests; \textit{data-centric security}, for trusting the data independently from where it came from.

In our opinion, such ICN characteristics could be effectively adopted in distributed NoSQL databases to \textit{simplify software development}, by exploiting the name-based API and  routing-by-name, \textit{improve performance}, by taking advantage of in-network caching and multicasting, provide \textit{data-level access control}, by using data-centric security.
%
Therefore, the contributions of this paper are: 
\begin{itemize}
\item a methodology for developing NoSQL distributed databases over Information Centric Networks; 
\item the practical application of the methodology for devising OpenGeoBase (OGB), a distributed spatial ICN/NoSQL database \cite{ogb}, now extended of temporal features; 
\item a new OGB performance evaluation.      
\end{itemize} 
 
To the best of our knowledge this is the first paper that propose to use ICN services for distributed NoSQL databases. In our previous work \cite{ogb}, we have described the use of ICN only for the specific case of a NoSQL \textit{spatial} database. In this paper we generalize the discussion, by proposing a methodology valid for \textit{generic} NoSQL databases.

\section{Related works}
\label{s:rw}
\begin{figure}[t]
	\centering
	\includegraphics[scale=0.23]{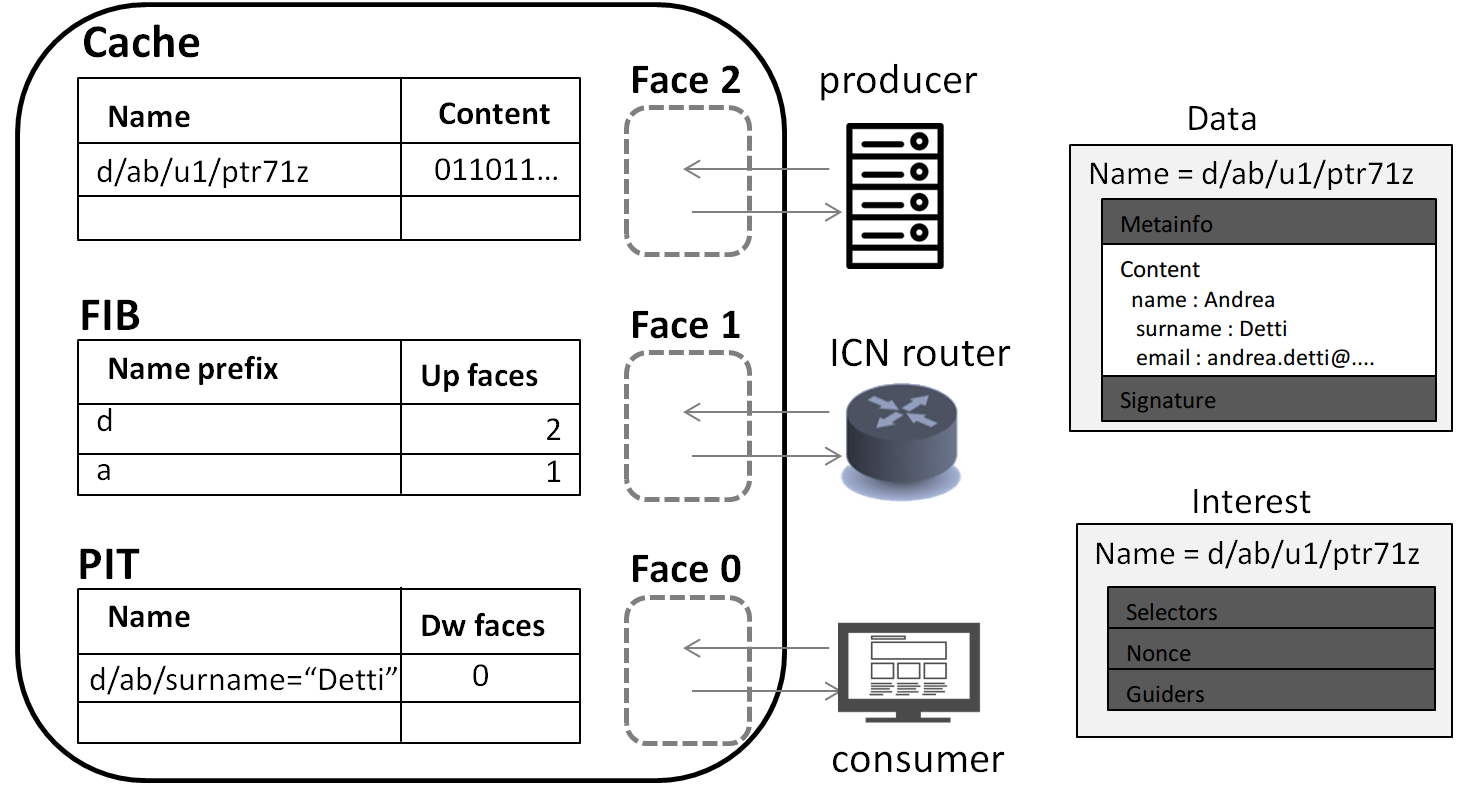}
	\caption{ICN forwarding engine model and packets}
	\label{f:icn-node}
	\label{f:aggregate}
\end{figure}

In what follows we provide the reader with minimal background on ICN and NoSQL databases. 

\subsection*{Information Centric Networks}
An ICN is formed by nodes that can be logically classified as consumers, producers and routers. Consumers pull named objects provided by producers, possibly going through intermediate routers. Any node uses the forwarding engine shown in fig. \ref{f:icn-node} and is connected to other nodes through channels, called \textit{faces}, which can be based on different transport technologies such as Ethernet, TCP/IP sockets, etc.. 

Data units exchanged in an ICN are called Interest and Data packets. To download a named object, a consumer issues an Interest packet that includes the object name (e.g. 'd/ptr71z') and that is forwarded upstream towards the producer. The forwarding process is a routing-by-name, i.e. a name-based prefix matching based on a Forwarding Information Base (FIB) containing name prefixes, such as 'd' and 'a' in case of fig.\ref{f:icn-node}. The FIB may be configured by routing protocols, where nodes advertise name prefixes rather than IP networks \cite{hoque2013nisr}. During the Interest forwarding, the engine temporary keeps track of the forwarded Interest in a Pending Information Table (PIT), by storing the name of the requested object and the identifiers of the downstream faces from which the Interest comes from. 

When an Interest reaches a node (producer or intermediate router) having the requested named object, the node sends back the object within a Data packet, whose header includes the object name. The Data packet is forwarded downstream to the consumer by consuming the information previously left in the PITs, as bread crumbs.

The forwarding engine caches (in-network) forwarded Data packets in a local store and immediately replies to Interest for cached Data. The data freshness is loosely controlled by an expiry approach. Any Data packet includes a freshness period metadata, chosen by the user, which indicates how long the engine should wait after the arrival of the Data before marking it as stale.

The forwarding engine also supports \textit{multicast} distribution. In case of concurrent Interests for a same object, the engine forwards only the first one, stores in the PIT the identifier of all arrival downstream faces and, when receives back the Data packet, relays a copy of it towards each downstream face. 

ICN is built on the notion of \textit{data-centric security}, for which the content itself is made secure rather than the connections over which it travels. ICN security framework provides each user with a private key and a ICN digital certificate, signed by a trust anchor, and uniquely identified by a name called \textit{key-locator} \cite{ndntrust}. Each Data packet is digitally signed by the content owner and includes the key-locator of the digital certificate to be used for signature verification. For access control purposes, Interest packets can be signed too.
 
Currently, different ICN implementation exists, the mostly used are NDN \cite{ndn} and CCNx \cite{ccnx} . 

\subsection*{NoSQL databases}

Database management systems (DBMS) may be based either on a relational model, or on a non-relational model also referred as NoSQL. 
For large information sets NoSQL databases are more and more replacing relational ones, since they can be easily distributed over different servers, and this feature fit well with cloud environments where databases are usually deployed. MongoDB, Cassandra, DocumentDB, etc. are popular NoSQL databases.

Many NoSQL databases are based on an \textit{aggregate-oriented} data model. An \textit{aggregate} is an object containing a set of information (a document, a row of related data, etc.), with an unique object identifier (oid) and seen as a self-consistent unit for DB operations, thus simplifying data distributions. The storage space of a database is logically organized in \textit{data sets}, i.e. group of related objects such as a MongoDB/DocumentDB collection, or a Cassandra columns-family. Furthermore, the storage space can be physically distributed over different servers (sharding). 

Temporal and spatial database functionality are usually integrated as extension/plug-in of a general purpose DBMS, which implement indexing procedures, such as R-Tree or Grid \cite{fox2013spatio}, and spatial/temporal query schemes, such as range queries, proximity queries, etc. However, although there exist numerous databases with spatial or temporal extensions (InfluxDB, Riak TS), databases including both extensions are rather limited, even though they are deemed to be very useful since often time and space exists together.

\begin{figure}[t!]
	\centering
	\includegraphics[scale=0.35]{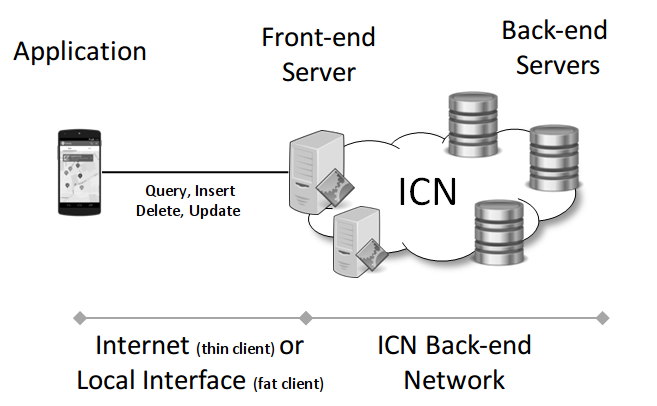}
	\caption{NoSQL/ICN distributed database architecture}
	\label{f:arch}
\end{figure}

\section{ICN/NoSQL distributed databases}
Figure \ref{f:arch} depicts the architecture of the proposed ICN/NoSQL database. It is formed by a cluster of servers working on top of an ICN layer rather than a TCP/IP one. The front-end servers are ICN consumers, and the back-end servers are ICN producers. Applications interact with front-end servers that exposes a typical NoSQL interface for querying, inserting and deleting objects. A front-end server could be co-located with the application (fat client solution) or running on a remote Internet device (thin client solution).

\begin{figure}[t!]
	\centering
	\includegraphics[scale=0.35]{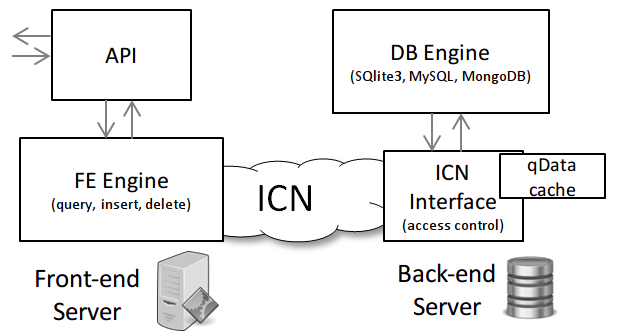}
	\caption{Front-end and back-end server functionality}
	\label{f:fun}
\end{figure}

Figure \ref{f:fun} shows the functional decomposition of the proposed architecture: front-end servers offer an Application Programming Interface (API) and an internal engine that satisfies client requests by using the ICN based procedures, hereafter described. The back-end servers are composed by an ICN interface that deals with ICN packets received-from or going-to the front-end servers, and by a local database engine that handles the local storage space. 

\subsection{Data model}
We readily observe that the NoSQL aggregate-oriented model perfectly fits the ICN paradigm simply considering an aggregate as an ICN named object, whose name is the object identifier (oid). Database objects can be requested by oid using Interest packets and received within Data packets. We dubbed these packets as \textit{oInterest} and \textit{oData}, for distinguishing them from other Interest/Data packets used for different purposes.  


\subsection{Data sharding}

Sharding is a method for distributing objects across multiple back-end servers. It uses an object attribute (\textit{sharding key}) and a \textit{sharding logic} to map the object onto a \textit{sharding domain}, which is partitioned in subsets called \textit{shards}. Each back-end server is configured for storing the objects belonging to one or more shards. 

\begin{figure}[t]
	\centering
	\includegraphics[scale=0.29]{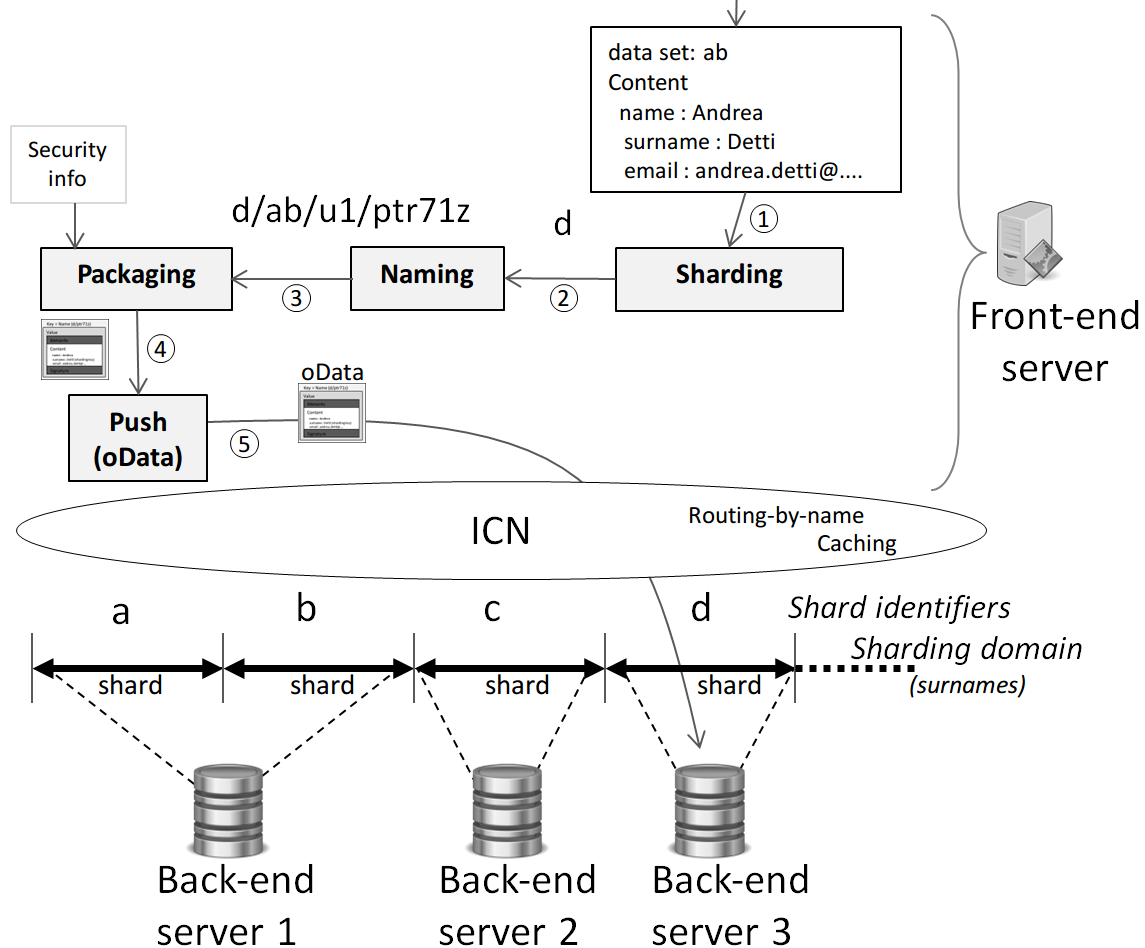}
	\caption{Data sharding and insert}
	\label{f:sharding}
\end{figure}

For instance, in fig. \ref{f:sharding} we consider a simple address book application in which the sharding domain is the set of possible surnames and each shard contains all the surnames starting with a given letter. The shards 'a' and 'b' are assigned to server 1, the shard 'c' to server 2, and so forth. The sharding key is the surname and the sharding logic simply extracts the first surname letter to identify the related shard. 

To ensure optimal performance and scalability the sharding strategy have to be selected in a way that is appropriate for the types of queries the application performs. When the sharding strategy is carefully designed, most of the queries can be sent only to those back-end servers that actually have the interested data (query routing), optimizing system load. Otherwise, the queries should be distributed to all the back-end servers (query flooding). For instance, the sharding strategy used in fig. \ref{f:sharding} allows query routing in case of queries finding by surname.  For other queries, flooding is necessary.

To support data sharding, the NoSQL/ICN database identifies each shard with a unique name, called \textit{shard identifier}. On the ICN routing plane, back-end servers advertise their shard identifiers as ICN name prefixes. Thus, ICN nodes can route-by-name Interest packets having the shard identifier as first name component towards the proper back-end server. 

For instance, in fig. \ref{f:sharding} the shard identifiers are 'a', 'b', 'c', 'd', etc.. Server 1 advertises the name prefixes 'a' and 'b', Sever 2 advertises the prefix 'c', and so forth.     


\subsection{Insert operation}
From the top down, fig. \ref{f:sharding} describes the object insert procedure of the NoSQL/ICN database. When a user inserts an object, a sharding logic computes the associated shard identifier on the base of the sharding key. In figure \ref{f:sharding} the sharding key is the surname 'Detti' and the computed shard identifier is 'd'. 

A naming function composes the object identifier as an ICN name, called \textit{oName}, by combining the shard identifier (\textit{sid}), the data set identifier (\textit{did}), the unique user identifier (\textit{uid}) and an unique application dependent suffix (e.g. a random string), as follows 

\vspace{5pt}
\noindent\texttt{\small oName = \{sid\}/\{did\}/\{uid\}/\{app suffix\}}
\vspace{5pt} 

\noindent In figure \ref{f:sharding} the sid is 'd', the did is 'ab' (Address Book), the uid is 'u1', and the application suffix is 'ptr71z'. 

Afterwards, a packaging function encodes the object as an oData packet and, finally, a Push procedure exploits ICN routing-by-name to deliver the oData packet to the responsible back-end server, which stores it in its local database. We observe that ICN natively provides pull services but not push ones; however in the literature there are different approaches for pushing data  \cite{franccois2013ccn} \cite{carzaniga2011content}, including our one in \cite{ogb}.


For some applications it may happens that an object "intersects" more than one shard. For instance, in the address book application example, if a contact has as two surnames, e.g. Blefari Melazzi, such an object belongs to two shards, 'b' and 'm'.  
We deal with a multiple shard object by packaging it in many oData packets, one for each shard, whose oNames differ for the shard identifier. We identify one of them as \textit{master} object, and the others as \textit{reference} objects. The reference objects contain the name of the master object, i.e. a references.  

\subsection{Query operation}

\begin{figure}[t]
 	\centering
 	\includegraphics[scale=0.29]{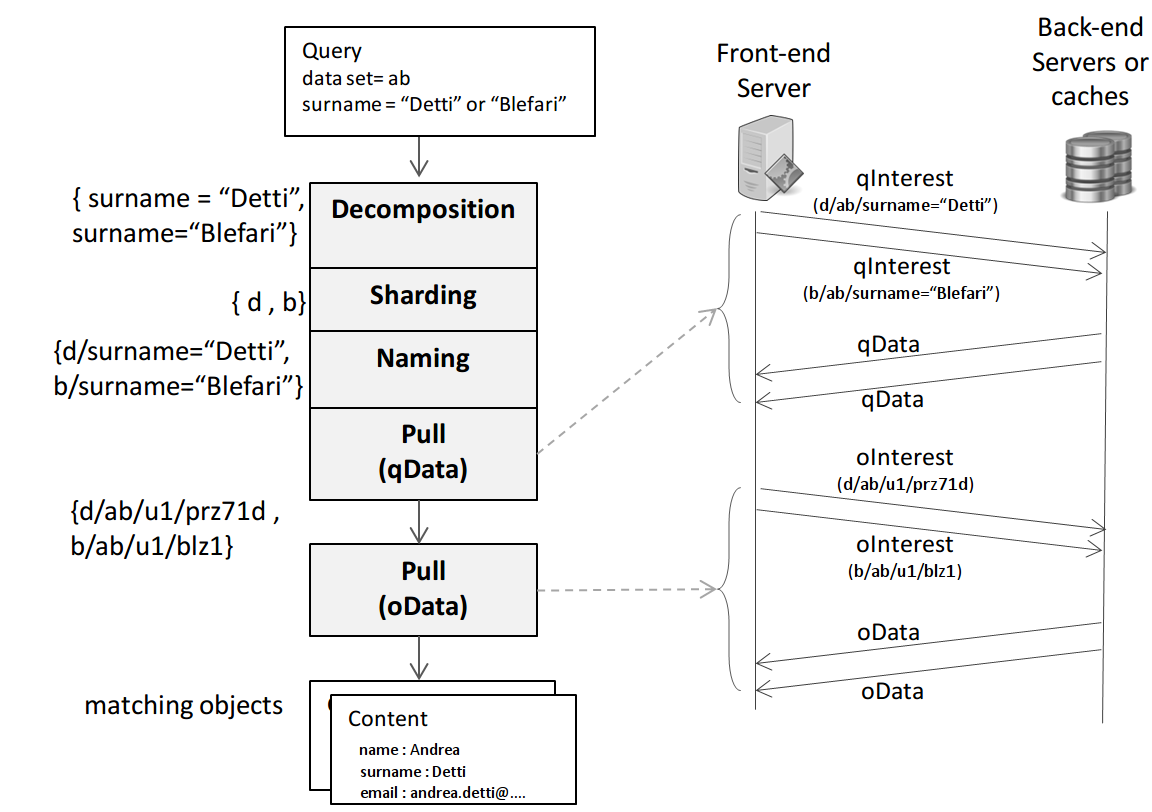}
 	\caption{Generic query procedure}
 	\label{f:qp}
 	\vspace{-5pt}
\end{figure}

A query is a request for database objects satisfying specific conditions. A query is solved by the front-end server as shown in fig. \ref{f:qp}. A sharding logic parses query parameters and computes the involved shards. For each shard, a naming function computes a name, called \textit{qName}, which is composed by the shard identifier (sid), the data set identifier (did) and query conditions, e.g. surname="Detti". 

\vspace{5pt}
\noindent\texttt{\small qName = \{sid\}/\{did\}/\{query conditions\}}
\vspace{5pt}

For each qName an Interest, called \textit{qInterests}, is sent out and it is routed-by-name by the ICN toward the appropriate back-end server. 
The receiving back-end server parses the qName to extract the query conditions, and then carries out a local query using the local database engine. 
The local query returns only the names (oNames) of the master objects that match the query conditions. The name list is packaged in a Data packet, called \textit{qData}, which is sent back to satisfy the qInterest. 

When all qData packets are received, the front-end server has a whole list oNames, which are then pulled through an oInterest-oData packet exchange. 
In so doing, we are actually solving a query in two \textit{pull} phases and this may sound as a temporal inefficiency. Anyway, we chose this approach both to transport only one time objects that intersect more shards, and to exploit the ICN in-network caching, as hereafter discussed.

\subsection{Caching}
Even though caching can dramatically accelerate query processing, its usage should be carefully designed in database applications, where it is likely not acceptable to send back stale data. For this reason we used two different caching strategies for qData and oData.

We observe that while the name of a qData remains unchanged, its content may change over time due to object insertions or removals. Indeed, the name embeds query conditions and the content contains the query result. As a consequence, we do not cache qData packets within the ICN forwarders (fig. \ref{f:icn-node}), since they use an expiry based freshness control and therefore stale data can be sent back. 
However, we deploy an application-layer qData cache within the back-end server, whose elements are immediately cleaned when object insertions or removals make them stale (fig. \ref{f:fun}). 

To increase the effectiveness of qData cache, the front-end server decomposes a complex query in a set of smaller sub-queries that may be requested more frequently than the complex one (fig. \ref{f:qp}). 
For instance, a query for ' surname="Detti" OR surname="Blefari" ' can be restructured as sub-query for surname="Detti" and a sub-query for surname="Blefari".  

For what concern the caching of oData packets, the cache inside the ICN forwarding engine can be safely used since their content never change. Moreover, when an object is removed, its oName will no more be included in any qData, thus the removed oData will be never fetched during the query procedure even if cached. 

\subsection{Delete operation}
To delete an object it is necessary to remove the master object and, in case, all its reference objects. The removal of an object is carried out issuing a command Interest, called \textit{dInterest}, that includes the oName of the object to remove followed by the "/DELETE" command string. A Data message called \textit{dData} containing the operation result is sent back.  

\subsection{Security}
The ICN/NoSQL database exploits data-centric security to achieve \textit{data-level access control} and \textit{data-level security}. The main idea is to encode the user access rights within the name (\textit{klName}) of its certificate, i.e. its key-locator. Namely, 

\vspace{5pt}
\noindent\texttt{\small klName = CERT/\{did\}/\{uid\}/\{permission[rw,r]\}}
\vspace{5pt}  

The data-level access control scheme provides that a user (uid) having a certificate for a given data set (did) is allowed to read (r) all the objects of such data set. Moreover, if the user has the writing permission (rw), she can also insert and remove objects in the data set. The ICN implementation of these policies requires to sign every Interest and Data packet sent by the front-end server and to verify the signature at the back-end side. Furthermore, a comparison between other identifiers is required as reported in table \ref{t:ac}. 

With regard to data security, the front-end server control the data integrity by verifying signature of Data packets. 

\begin{table}[]
	\centering
	\caption{Access control checks for Insertions (I), Queries (Q) and Deletions (D) }
	\label{t:ac}
	\begin{tabular}{lll}
		\hline
		op. & identifier checks & perm.\\ \hline
		I        & oName.did=klName.did AND oName.uid=klName.uid    &  rw \\ \hline
		Q        &  oName.did=klName.did, qName.did=klName.uid	 & r OR rw \\ \hline
		D        &  dName.did=klName.did AND dName.uid=klName.uid	& rw \\ \hline
	\end{tabular}
\vspace{-5pt}
\end{table}

\section{OpenGeoBase}
\label{s:ogb}
\subsection{Description}
OpenGeoBase (OGB) is a distributed NoSQL spatial database, whose current release provides also some temporal features. Users can store spatial objects, structured as GeoJSON Feature objects \cite{geojson}. In addition, each object could have a temporal extent, specifying the time period during which the information reported in the object is valid.
   
For instance, the following GeoJSON could be used by a city traffic monitoring application, where mobile sensors periodically insert information regarding sensor name and speed, GPS lng/lat, and measurement valid period in Unix epoc time.  

\vspace{3pt}
\noindent\texttt{\small
	\{"type": "Feature", \\
	"geometry": \{"type": "Point","coordinates": [12.51133, 41.8919]\},\\
	"properties": \{"sensor-name": "sensor1", "speed" : "23 km/h" \},\\
	"temporalExtent": \{"validTime": \{"type":"interval","value": [24807931 24807931]\}\}
	}

\vspace{3pt} 

OGB users can carry out either inclusion or intersect spatial range queries for obtaining all the GeoJSON objects completely (inclusion) or partially (intersect) contained in the range area, i.e. a 2D bounding box. Moreover, range queries could be time bounded, i.e. limited only to objects valid in a given time interval.

The sharing strategy of OGB considers the physical space as the sharding domain. The domain is partitioned in shards that are squared geographical zones, aligned with the lng/lat GPS grid, and whose side is 1 degree long. The shard identifier (sid) has the form \texttt{\small lng/lat}, where lng/lat is the south-west coordinate of the zone.
         
To optimize qData caching, a query decomposition strategy (fig. \ref{f:qp}) tessellates the query area with a limited number of covering tiles (e.g. 40), where tiles can have a side length of 1 or 0.1 or 0.01 degrees. Furthermore, in case of a spatio-temporal query there is an additional decomposition that divides the requested time interval in a limited number of covering periods (up to 5), whose size may be 1, 10, 100,1000 or 10000 minutes. For each couple tile/period a sub-query is carried out, and then returning GeoJSON objects are collected and sent back to the application. 

OGB is implemented using a thin client approach (fig. \ref{f:arch}); the front-end server logic runs within a Spring STS Application Server; the ICN is based on NDN software \cite{ndn}; the back-end server is a modified version of ndn-repo, which uses SQLite3 as local database engine.

\subsection{Performance evaluation} 
\begin{figure*}[ht]
	\centering
	
	\subfigure[Maximum query rate vs query area, GTFS data set, Poisson inter-arrivals]{  
		\includegraphics[scale=0.53]{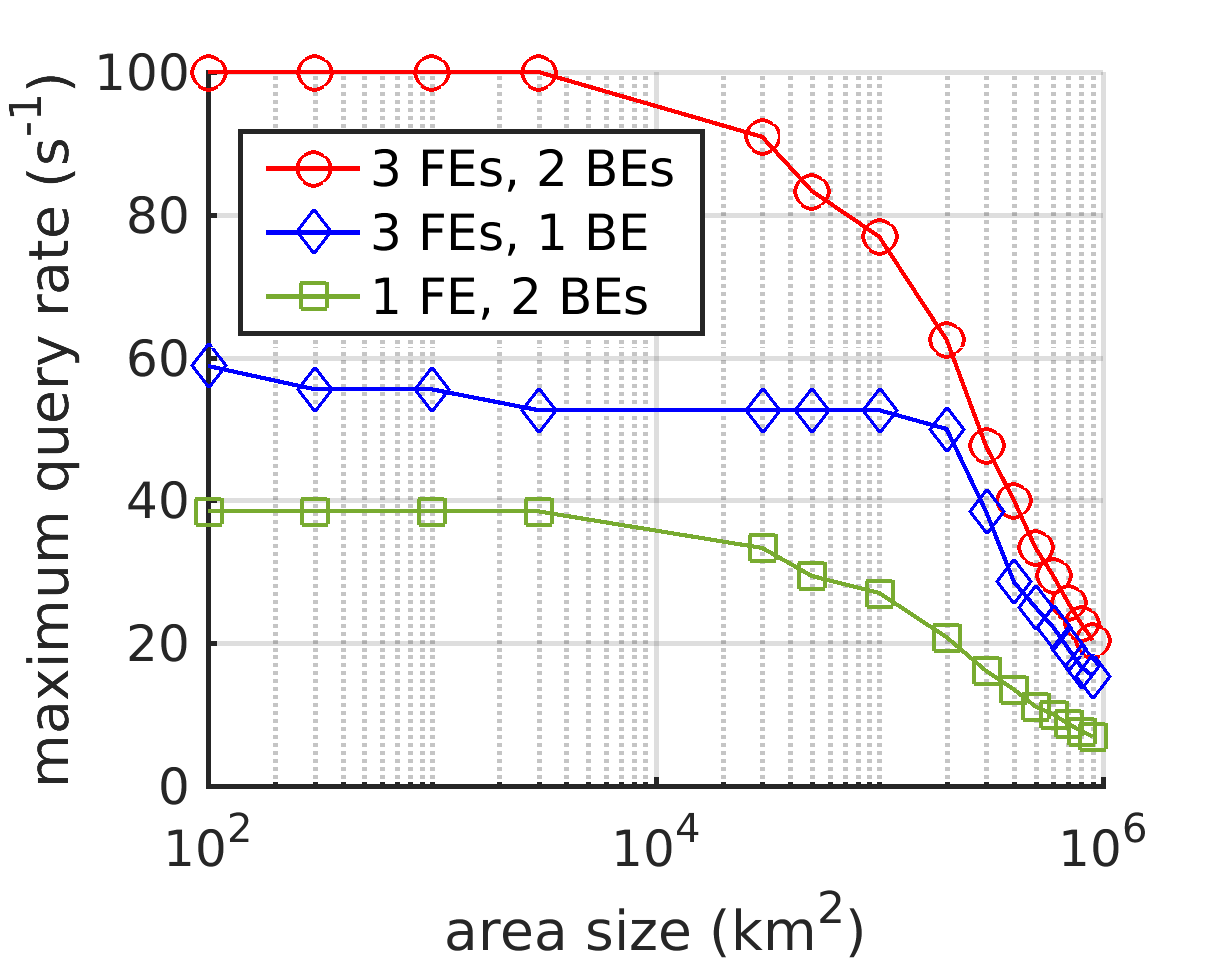}
		\label{f:GTFSscal}
	}\hfill
	\subfigure[Query delay vs query number, GTFS data set, Poisson inter-arrivals, query area $10^5 $km$^2$, 1 FE, 2 BE]{  
		\includegraphics[scale=0.53]{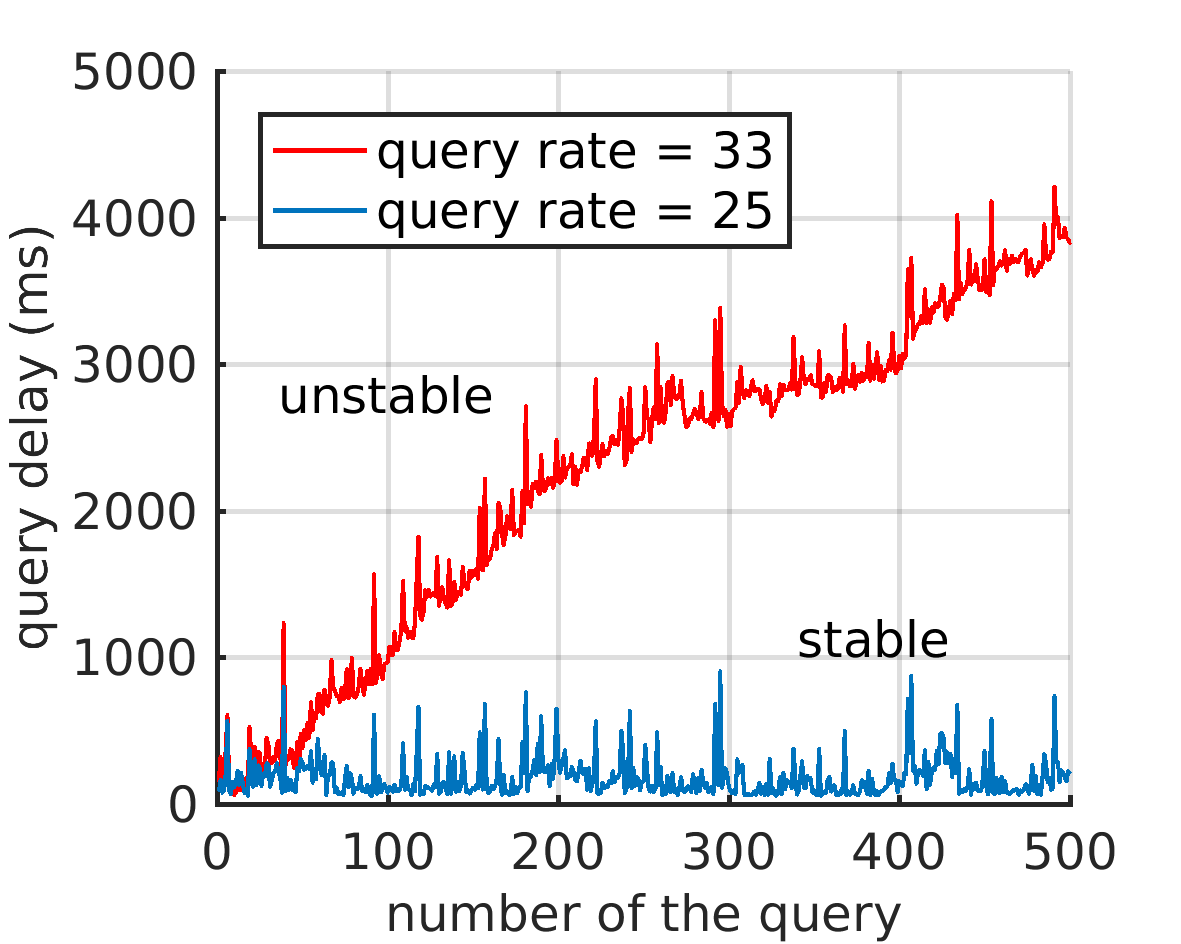}
		\label{f:GTFSstab}
	}\hfill
	\subfigure[Average query delay vs query area, GTFS data set, Poisson inter-arrivals, query rate equal to the half of maximum rate (unloaded condition)]{  
		\includegraphics[scale=0.53]{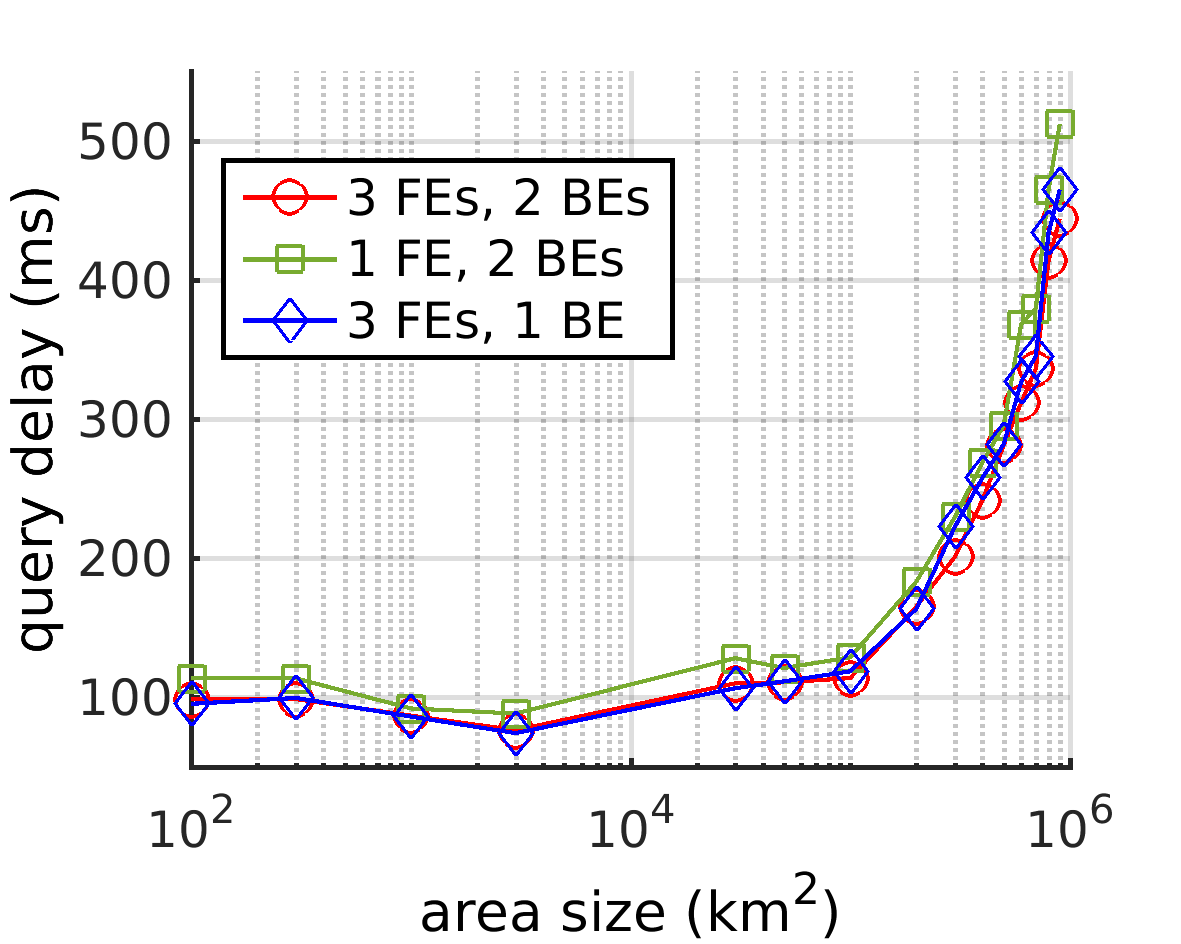}
		\label{f:GTFSunload}
	}\hfill
	\subfigure[Average query delay vs query rate, GTFS data set, Poisson inter-arrivals, query area $10^5 $km$^2$]{  
		\includegraphics[scale=0.53]{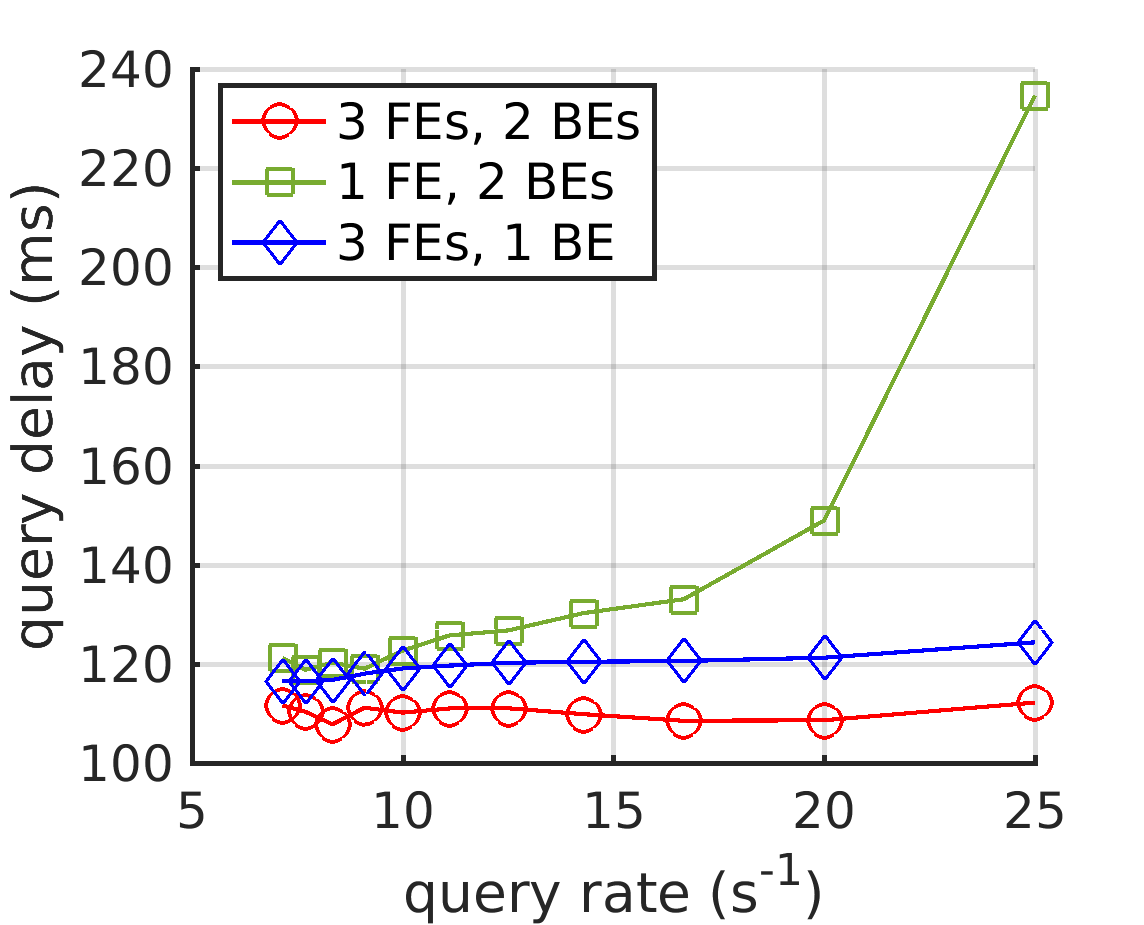}
		\label{f:GTFSqdel}
	}\hfill
	\subfigure[Maximum query rate vs temporal query size, Rome bus data set, Poisson inter-arrivals, query area 1 km$^2$ ]{  
		\includegraphics[scale=0.53]{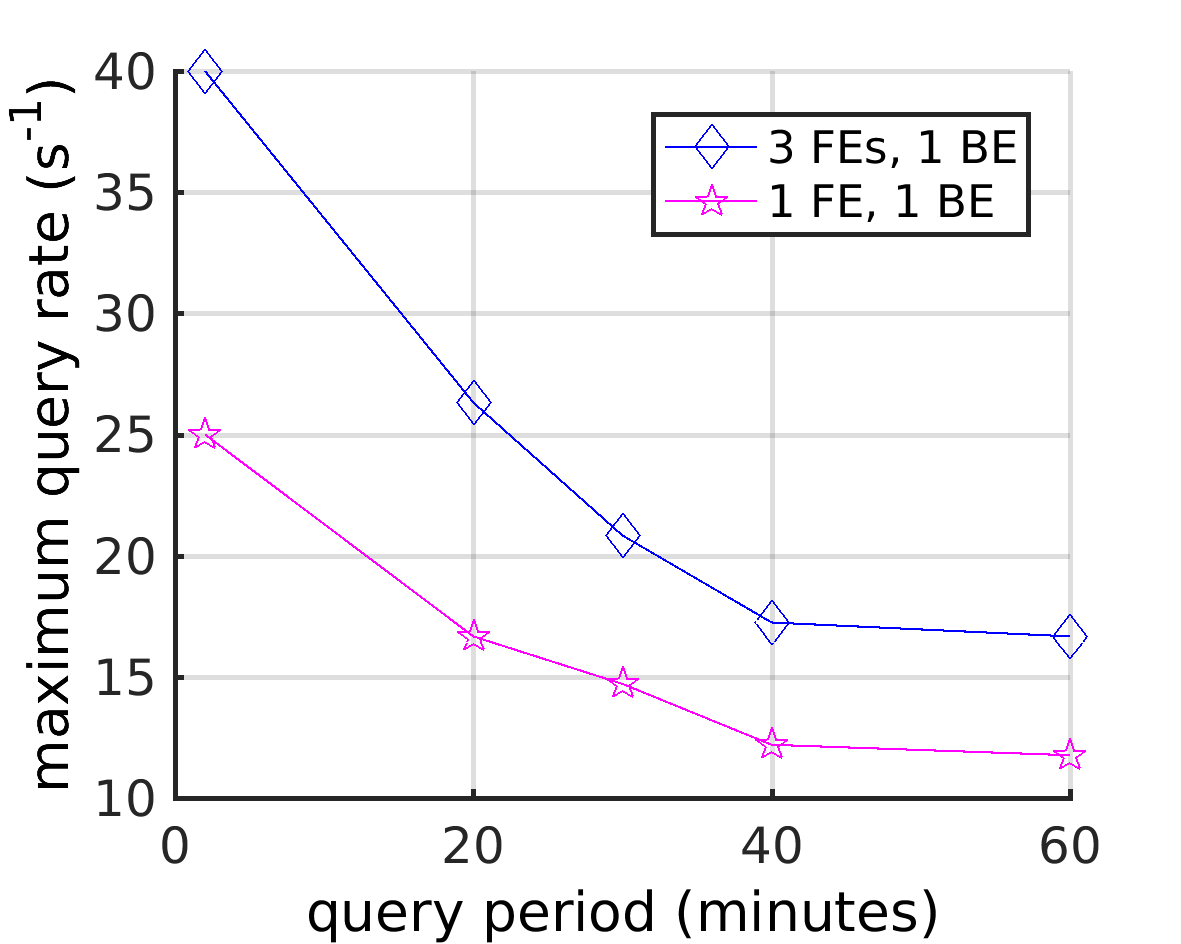}
		\label{f:BusScal}
	}\hfill
	\subfigure[Maximum insert rate vs number of FE/BE servers, objects with point geometry, Poisson inter-arrivals, signed (default) and unsigned objects]{  
		\includegraphics[scale=0.52]{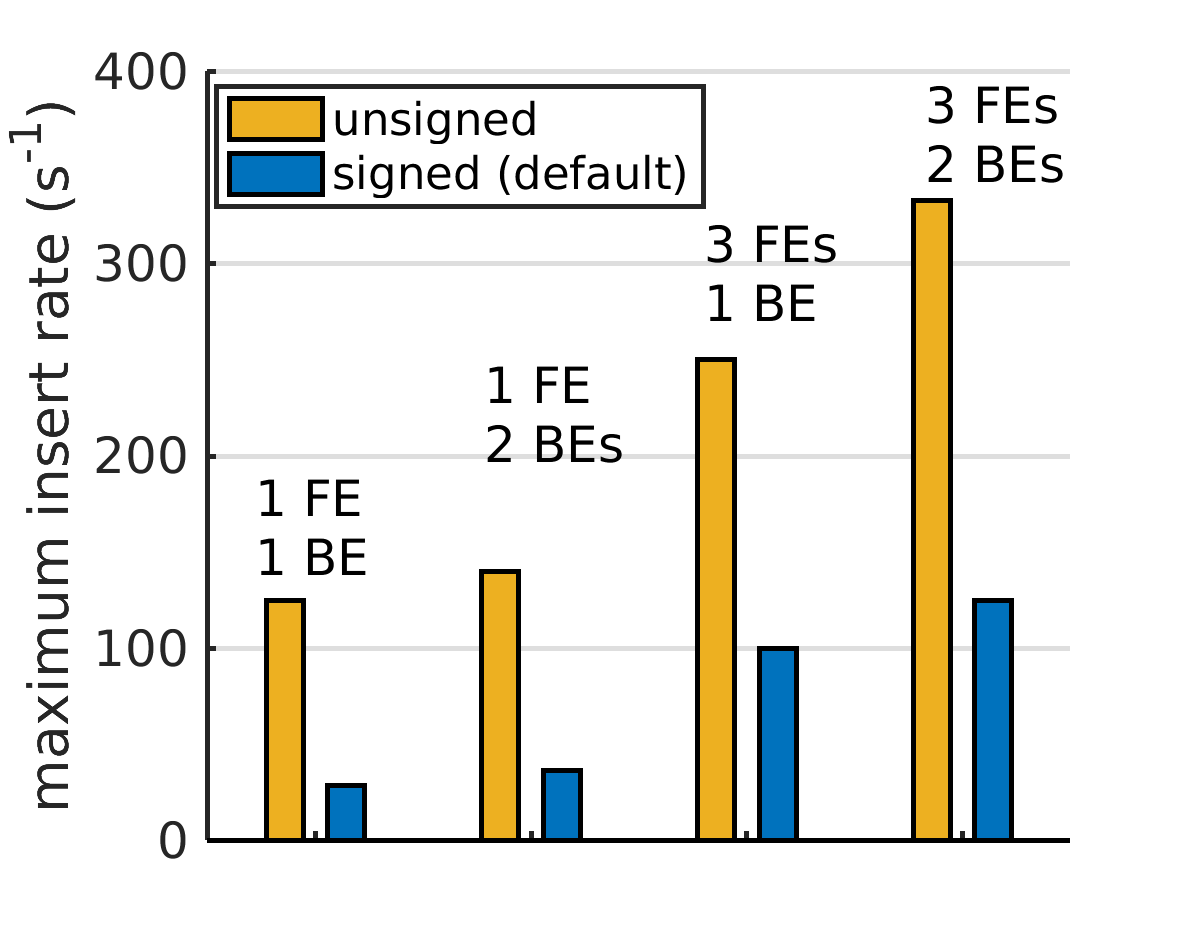}
		\label{f:InsertScal}
	}\hfill
	\caption{Performance evaluation results}
	\vspace{-10pt}
\end{figure*}     
We carried out a performance evaluation using two European real data sets, related to Intelligent Transport System (ITS) applications \cite{webapp}. The first data set, named "GTFS", contains information regarding European public transports such as: stop coordinates, schedules, time-tables, etc. This information has been derived by downloading a thousand of public GTFS files from Internet. For each GTFS file, we inserted a GeoJSON multipoint object, without temporal extension, where each point is associated to a stop. Several stored objects are composed by thousands of points (e.g. all train stops of a country) and their size may even reach the MByte order.       

The second data set, named "Rome bus", is spatio-temporal and contains the positions of 2870 buses in Rome, sampled with a step of 5 minutes during a single day. Each bus measurement is stored as a GeoJSON point object, and includes bus detail, position and sample time.

The considered cluster architecture is formed by a set of front-end (FE) servers, a set of back-end (BE) servers, and a benchmark application that uniformly distributes queries and insertions among available FE servers. These software components run on different virtual machines, connected each other by a Linux bridge.     

Fig. \ref{f:GTFSscal} reports the maximum range query rate for the GTFS data set versus the range query area, for different cluster configurations, i.e. number of FE and BE servers. 
Maximum rate is the highest rate for which the time needed to solve a query (query delay) has a stable behavior versus time, as shown in fig. \ref{f:GTFSstab}. It has been measured by loading the system with a sequence of range queries, randomly located in Europe, and whose inter-arrival time follows a Poisson distribution. 

The cluster with the greatest resources, 3 FE and 2 BE servers, supports the highest rate, thus confirming the database capability to horizontally scale. Performance improves both increasing the number of BE servers, thanks to data sharding and query routing, and increasing the number of FE servers, thanks to load balancing. The maximum rate decreases increasing the range query area,  since queries require more processing and data transfer. 

Fig. \ref{f:GTFSunload} shows the average query delay versus the area size, in case of an unloaded system. Unloading conditions are reproduced submitting range queries with a rate equal to the half of the maximum one. There is no practical difference among the cluster deployments since increasing resources, i.e. FE or BE servers, is only needed in overloading conditions. We point out that the maximum rates reported in fig. \ref{f:GTFSstab} are much greater than the inverse of query delays reported in \ref{f:GTFSunload}. However, this result is not  surprising since it is a consequence of our multi-threads implementation of the FE server, whereby client requests can be served in parallel.            

Fig. \ref{f:GTFSqdel} reports the average query delay versus the query rate in case of range queries of $10^5$ km$^2$. From the comparison between 3FE-1BE and 1FE-2BE we infer that for OGB it is more effective to increase the number of FEs than the number of BEs. In facts, the processing load of a FE server is greater than the one of a BE server. It is worth to note that delays of 1FE-2BE deployment rapidly grows up since we are reaching its maximum sustainable rate. 

Fig. \ref{f:BusScal} shows the maximum query rate in case of spatio/temporal queries carried out on the Rome bus data set. We consider range queries whose spatial extension is 1 km$^2$, randomly located in Rome, and requesting objects that are valid in a given period.  Being Rome fully contained in a single shard (12 lat, 42 lng), we used a single back-end server and only changed the number of front-end servers. Also in this case having more resources makes it possible to sustain a greater rate. Increasing the query temporal extent, query rate decreases since more objects are sent back, and also the front-end decomposition processing is higher.    

Fig. \ref{f:InsertScal} reports the maximum insert rate measured by storing spatio/temporal objects with point geometry. Also for insertions, performance improves scaling out the cluster resources. We remark that differently from traditional spatio/temporal databases, in our case each query or insertion requires a digital signature verification and/or computation that adds few milliseconds of processing delay but provides data-level access control feature. To give an idea of the security impact, in fig. \ref{f:InsertScal} we also reported a case in which inserted objects are unsigned.

\section{Conclusions}
We exploited ICN technology to develop NoSQL distributed databases. The results obtained from a practical implementation, with real data sets, have shown the ability of ICN/NoSQL databases to effectively support horizontal scalability, caching, data sharding and data level access control.


\bibliographystyle{IEEEtran}
\bibliography{paper}

\clearpage

\ifpaper
\else	
\fi

\end{document}